# Transfer of Vertical Graphene Nanosheets onto Flexible Substrates towards Supercapacitor Application

Gopinath Sahoo[1,2†], Subrata Ghosh[2]*[†], S. R. Polaki[1,2]*, Tom Mathews[1,2] and M. Kamruddin[1,2]

[1] Surface and Nanoscience Division, Materials Science Group, Indira Gandhi Centre for Atomic Research-Homi Bhabha National Institute, Kalpakkam, India – 603102

[2] Surface and Nanoscience Division, Materials Science Group, Indira Gandhi Centre for Atomic Research, Kalpakkam, India – 603102

**Abstract**

Vertical graphene nanosheets (VGNs) are the material of choice for next-generation electronic device applications. The growing demand for flexible devices in electronic industry brings in restriction on growth temperature of the material of interest. However, VGNs with better structural quality is usually achieved at high growth temperatures. The difficulty associated with the direct growth on flexible substrates can overcome by adopting an effective strategy of transferring the well grown VGNs onto arbitrary flexible substrates through soft chemistry route. Hence, we demonstrated a simple, inexpensive and scalable technique for the transfer of VGNs onto arbitrary substrates without disrupting its morphology and structural properties. After transfer, the morphology, chemical structure and electronic properties are analyzed by scanning electron microscopy, Raman spectroscopy and four probe resistive methods, respectively. Associated characterization investigation indicates the retention of morphological, structural and electrical properties of transferred VGNs compared to as-grown one. Furthermore the storage capacity of the VGN's transferred onto flexible substrates is also examined. A very lower sheet resistance of 0.67 kΩ/□ and excellent supercapacitance of 158 μF/cm$^2$ with 91.4% retention after 2000 cycles confirms the great prospective of this damage-free transfer approach of VGNs for flexible nanoelectronic device applications.

Corresponding authors: subrataghosh.phys@gmail.com (Subrata Ghosh);
poalki@igcar.gov.in (S. R. Polaki)

[†] both authors equally contributed to the work

## 1. Introduction

Vertical Graphene Nanosheets (VGNs), one among the remarkable graphene family, emerges as potential candidate for the wide range of applications in the field from energy storage to bio-applications, to sensor, field emission, spintronic, optoelectronic right from its discovery.[1-5] The VGNs are a three-dimensional porous network made of vertically standing few layer graphene sheets. Each sheet composed of 2-10 number of graphene layers with few microns in length and height. The fascinating properties of VGNs includes high surface to volume ratio, open porous network, sharp edges, excellent electronic and thermal conductivity, thermal and chemical stability.[6-9]

Plasma Enhanced Chemical Vapor Deposition (PECVD) is recognized as one of the suitable technique to synthesis VGNs because of inherent electric field, faster growth rate, low temperature and catalyst-free growth over large area.[8] Depending upon the requirement, VGNs can be grown on any arbitrary substrate. However, the nucleation, growth rate and structural quality of VGNs are found to have significant dependency on the substrate.[10] In addition, thermal and hot-filament CVD are also employed to fabricate this novel nanoarchitecture.[3, 11] However, the VGNs are used to grown at high temperature, which limits the choice of substrates. The kind of substrate to be used for growth is defined by the application. Hence, the demand for utilization of VGNs on flexible substrate needs the growth at lower temperature such that the advantages of VGNs can be properly used for flexible electronic device applications. Till now, the lowest growth temperature of 225 °C for the VGNs growth is demonstrated by Park *et al.*[12] Moreover, higher degree of graphitization and morphology with more open space, which is beneficial for faster charge transfer kinetics as an electrode material, are not achievable at lower growth temperatures. In this context of recent scientific and technological advances in VGNs applications, the only convenient strategy is to subsequent transfer of the film on any arbitrary substrates without sacrificing its structural quality and morphology. With the exciting potential towards suitable applications, few groups attempted to tackle the challenging issue of VGNs transfer.[13, 14] Two-step transfer process of VGNs on arbitrary substrate is successfully developed by Quinlan *et al.*: (i) spin coating of polystyrene solution on VGNs, where quality is depends on spin-rate and (ii) removal of polystyrene using different solvents (MEK and NMP) [13]. The process involves several chemical processes, which can degrade the structural quality. Consequently, an alternative approach is being sought by Constantinescu *et al.* through novel laser-induced transfer technique (LIFT) to transfer VGNs onto soft electrode.[15] However, LIFT has the major negative aspects in preserving the morphology and it needs special handling skills. Recently, Yamada *et al.* investigated the optical properties of VGNs by transferring it from Cu substrate to quartz. However, morphological quality, structural quality, electrical properties and electrochemical stability of transferred film are not discussed in detail.[16] This background offers an opportunity to look for an alternative, scalable and simple technique to transfer VGNs on arbitrary substrates.

Herein, we present the simple one-step transfer technique of VGNs onto flexible substrate. We grown VGNs at 800°C to ensure higher structural quality and transfer them onto flexible substrate through simple chemical route. We characterized the transferred-VGNs by scanning electron microscope, Raman spectroscopy and four-point probe method to illustrate the compatibility of this method for desired device applications. By facilitating this feasible transfer approach, electrochemical investigation of transferred VGNs as supercapacitor electrode is carried out to demonstrate its versatility for flexible and foldable energy storage devices.

## 2. Experimental methods

### 2.1. Growth of Vertical Graphene Nanosheets (VGNs)

VGNs were grown on Ni substrate using ECR-PECVD system. Ultra high pure Ar and $CH_4$ were used as dilution and source gas, respectively. Prior to the growth, the chamber was evacuated to $10^{-6}$ mbar by turbo molecular pump backed by a rotary pump. Thereafter, substrate was subjected to annealing at 800°C and subsequently cleaned by Ar plasma with 200W power. The selection of gas composition and their ratio were carried out from our previous investigation.[17] The hydrocarbon gas ($CH_4$) of 5 sccm along with Ar is fed into the chamber for 90 min at same temperature. Growth of VGNS was carried out with microwave power of 375 W, at an operating pressure of $2\times10^{-3}$ mbar. The grown film was further subjected to post growth annealing for 30 min at 800 °C by just switching off the microwave plasma. Finally, the sample was allowed to cool down to room temperature and taken out from the vacuum chamber for further characterization.

## 2.2. Transfer process of VGNs

As-grown VGNs on Ni foil is transferred onto flexible substrates. Our process is similar with the CVD grown graphene transfer process, as reported by Kim *et al.* [18] The VGNs were transferred onto variety of substrates includes glass slide, Parafin film, Over Head Projector (OHP) sheet, paper, carbon cloth etc. The large area transfer process involves several steps as follows. Prior to the transfer, as-grown VGNs/Ni sample was annealed at temperature of 80 °C for 10 min. In order to etch the Ni-foil, VGNs/Ni is immersed in 1M $FeCl_3$ solution for 90 min. The ionic etching equation without forming gaseous species as follows: [18]

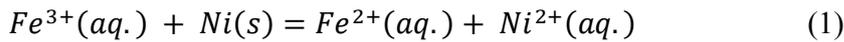

$$Fe^{3+}(aq.) + Ni(s) = Fe^{2+}(aq.) + Ni^{2+}(aq.) \qquad (1)$$

At the end of the process freely floating VGN film on $FeCl_3$ solution is evident. It confirms the complete etching of Ni substrate. Before transferring the VGN film onto a desired substrate, the residual $FeCl_3$ was removed from petri dish using a syringe upto certain level and dilute the solution further by pouring equal amount of DI water. This procedure of dilution with DI water was repeated 4 to 5 times with utmost care because any disturbance to the freely floating VGN film can break it into pieces. Prior to transfer process, the desired substrate was separately cleaned by sonicating in isopropyl alcohol and dried in nitrogen gas. Subsequently, the floated film was transferred onto the desired substrate by fishing it. The transferred film was baked for 2 min at 50 °C to get rid of trapped water at the interface of substrate and VGNs, which can lead to cracks. The transfer process was completed within 2 hrs. The whole procedure was tried for many times to ensure the reproducibility. Noteworthy, the transferred procedure does not need polymer support and its removal step afterwards like the transfer process introduced by Quinlan *et al.* [13] The advantageous of this procedure is very simple, high reproducibility and large throughput. Note that, during the fishing, one needs to be extremely careful to avoid cracking of the sample and breaking down into several pieces.

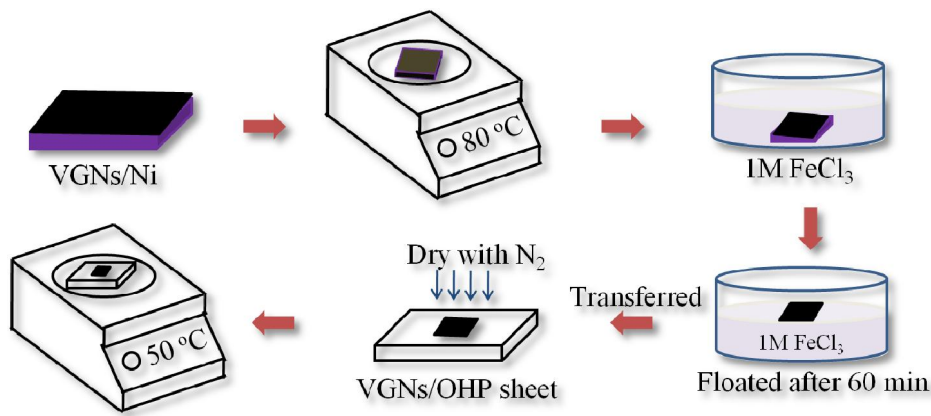

Fig. 1: Schematic representation of transfer process for vertical graphene nanosheets onto flexible substrate

## 2.3. Characterization

The surface morphology of both as-grown and transferred VGNs was examined by Field-emission Scanning Electron Microscopy (Supra 55, Carl Zeiss, Germany). Raman spectroscopic investigation was carried out in order to quantify the crystalline quality of as-grown and transferred VGNs using micro-Raman spectroscopy (in-Via Ranishaw, UK). It was also employed to find out the structural modification introduced by the transferring process. Raman spectra were recorded in 1000-3500 $cm^{-1}$ frequency region using 514 nm laser with 50× objective lens (Numerical Aperture of 0.8). The electrical measurement of the sample was carried out using Agilent B2902A precision source/measure unit in four-probe resistive method. Ag paste was used to provide electrical contact. Electrochemical investigation of transferred VGNs for supercapacitor application was carried out in three electrode system using Metrohm-Autolab electrochemical work station (model PGSTAT302N, Netherland). The current-voltage behaviors were recorded in a potential window of 0.5 V vs Ag/AgCl (3M KCl saturated) at scan rates ranging from 100 to 500 mV/s. Pt foil was used as counter electrode and 1M KOH was chosen as electrolyte for this study.

## 3. Results and discussions

Figure 1 depicts the uniform and wrinkle free transfer of VGNs onto the flexible substrate over the area of $1 \times 1$ $cm^2$. It is also evident from fig 2 that the transferred VGN film is well adherent and flexible. The scanning electron micrographs of as-grown and transferred VGNs are shown in Fig. 3 (a) and (b), respectively. The SEM micrographs confirm that no visual change in structure and morphology. The non-agglomerated and porous network of VGNs architect is evident from scanning electron micrograph, which is beneficial for providing multiple channel/pathways to access the electrolyte ions towards enhanced supercapacitor performance. Notably, we obtained uniform VGNs over a large area, without cracks, contaminations and deterioration. The observed results confirm the preservation of morphology after transferring the film by wet-chemistry method.

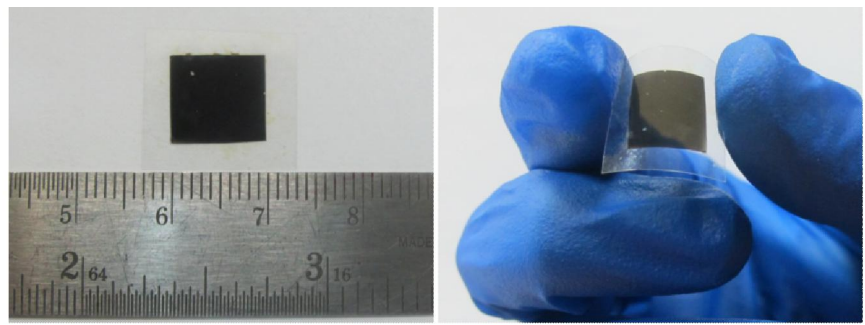

Fig. 2: O[...]ate

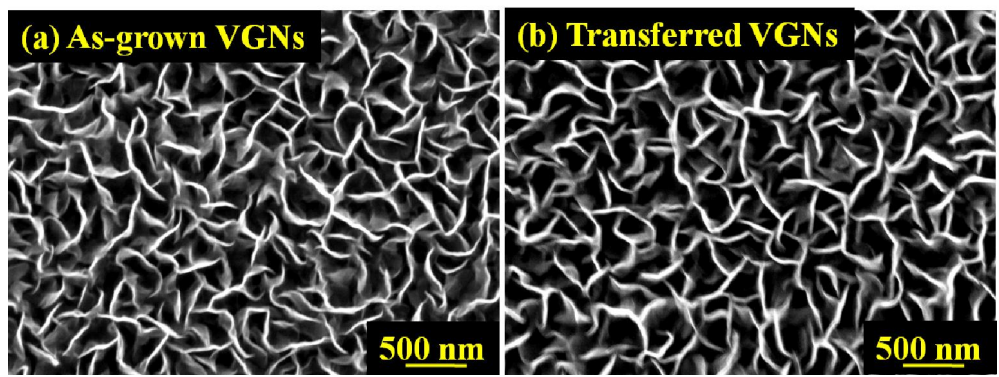

Fig. 3: Scanning Electron Micrographs of (a) as-grown and (b) transferred vertical graphene nanosheets

In order to confirm the structural integrity after transfer process, Raman spectroscopic analysis was carried out on both as-grown and transferred VGNs. Figure 4 shows the obtained Raman spectra associated with the as-grown and transferred VGNs. Typical Raman spectra of VGNs consists of three prominent peaks: D- bands (~1350 cm$^{-1}$), G-band (~1580 cm$^{-1}$) and G'-band (~2700 cm$^{-1}$).[19, 20] The D-band (TO phonon) corresponds to the breathing mode of hexagonal ring.[19, 20] The G- band associated with the iTO and LO phonon mode (E$_{2g}$ symmetry) and arises due to stretching mode of $sp^2$ bonded carbon in ring and chain.[19, 20] The G'-band is originated due to the two iTO phonon double resonance process.[19, 20] The pronounced D band and other defect related bands, D" band (1100 cm$^{-1}$), D' band (1620 cm$^{-1}$) and their overtones at 2456 cm$^{-1}$, 2945 cm$^{-1}$ and 3240 cm$^{-1}$ confirms the defected nature due to the high ion bombardment, huge edge density, heptagon-pentagon structure and presence of C-H related content.[20-22] Lower full width at half maximum (FWHM) of D-, G- and G'- band is an indicative for higher degree of crystallinity. The shift in peak position of those prominent peaks is the signature of strain, doping and disorder.[20] The first order and second order Raman spectra are deconvoluted by Lorentzian line shape. Insignificant change in normalized Raman spectra of as-grown and transferred VGNs confirms from the Fig. 4. The Raman spectra of VGNs are recorded at several places before and after the transfer process. The average FWHM of D-, G- and G'- band of as-grown VGNs are found to 38, 31 and 70 cm$^{-1}$, respectively. Whereas, measured average FWHM of D-, G- and G'- band of transferred VGNs are 39, 32 and 73 cm$^{-1}$, respectively. A very less shift in peak position and almost unchanged FWHM suggests that the crystalline quality is maintained even after transfer the VGNs on to a flexible substrate.

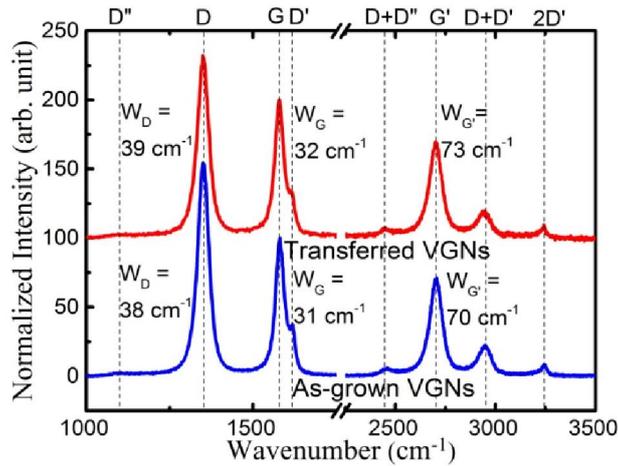

Fig. 4 : Raman spectra of as-grown and transferred vertical graphene nanosheets

The slight broadening in FWHM and shift in peak position are attributed to the substrate effect and trapped water into the interface between VGNs and flexible substrates. Therefore, inspection of the transferred film by Raman spectroscopy ensures that the procedure does not introduce any significant structural degradation and VGNs were uniformly transferred on to the target substrate over an area of 1 × 1 cm$^2$.

Apart from the structural quality, electrical conductivity of the materials plays major role in faster charge transfer kinetics for the potential utilization. In view of that, the electrical measurement was carried out for the transferred VGNs by four-point probe station assuming continuous thin film in macroscopic scale. The linear current-voltage relationship of VGNs from the both configuration ensures the Ohmic contact, Fig. 4. Te sheet resistance of VGNs is estimated from the van-der Pauw equation and found to be 0.67 kΩ/□. The lower sheet resistance is beneficial in faster charge transport as an electrode material.

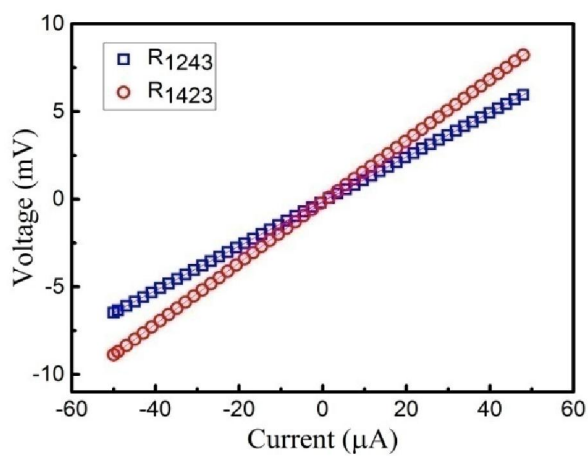

Fig. 4 : Current-Voltage relationship of transferred vertical graphene nanosheets

The successful transfer is also ensured by investigating the supercapacitative performance of VGNs by exposing it to electrolyte. For the purpose of supercapacitance study, VGNs of 2 × 1 cm$^2$ area is transformed onto flexible substrate to keep the electrical contact away from actual area of 1 × 1 cm$^2$ exposed to the electrolyte. We anticipated the possibility of transferring VGNs by this simple and scalable approach over large areas, which paves the way for industrial application of this nanoarchitect. The panel (a) of Fig. 5 represents the cyclic voltammogram (CV) of transferred VGNs, in KOH aqueous electrolyte, at different scan rate ranges from 100 mV/s to 500 mV/s. A near rectangular CV and its unaltered shape with scan rate confirm the excellent supercapcitor performance. The areal capacitance is calculated from the CV using the following equation

$$C_A = \frac{\int I\, dV}{\Delta V . A . s} \qquad (2)$$

where, $\int I\, dV$ represents area under the curve, $\Delta V$ is the potential window, $s$ is the scan rate and A is the electrode area exposed to the electrolyte ($1 \times 1$ cm$^2$). The areal capacitance is found to be 158 µF/cm$^2$ at a scan rate of 100 mV/s. The suprercapacitive performance can be improved further by fabricating the structure with higher inter-sheet distance, changing the wettability, using proper electrolyte, and encapsulating heteroatoms.[4, 23-27] However, the obtained capacitance value is in good agreement with the existing literatures.[1, 23] Figure 5(b) represents the areal capacitance with respect to the scan rate and found to be decreased with scan rate can be attributed to the inaccessibility of electrolyte ion to the interior of material.

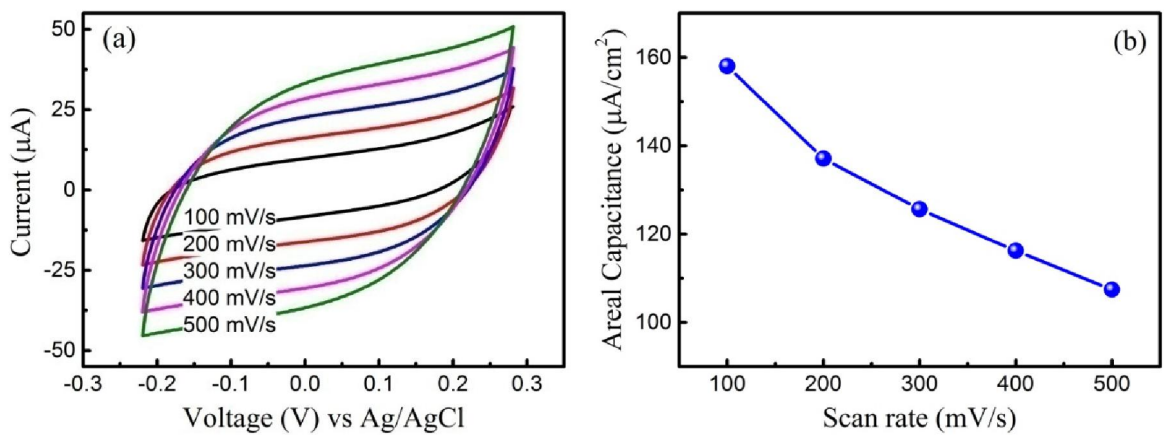

Fig. 5 : (a) Cyclic voltammogram (CV) of transferred VGNs in 1M KOH at scan rate 100 – 500 mV/s and (b) plot of Areal capacitance of transferred VGNs with respect to scan rate

The galvanostatic charge/discharge, is also performed, revealing the symmetric and near linear profile even at high current density of 20 µA/cm$^2$ with insignificant voltage drop.(Fig. 6(a)) The obtained result is also good indicative of ideal supercapacitor behavior. In addition, electrode materials must have adequate electrochemical stability during large number of charge-discharge cycle. Hence, charge/discharge experiment was carried out for 2000 cycles at a scan rate of 12 µA/cm² to probe its electrochemical stability and the result is depicted in Fig. 6(b). The supercapacitive retention of transferred VGNs is found to be 91.4 % even after 2000 charge/discharge cycles, as shown in Fig. 6(b).

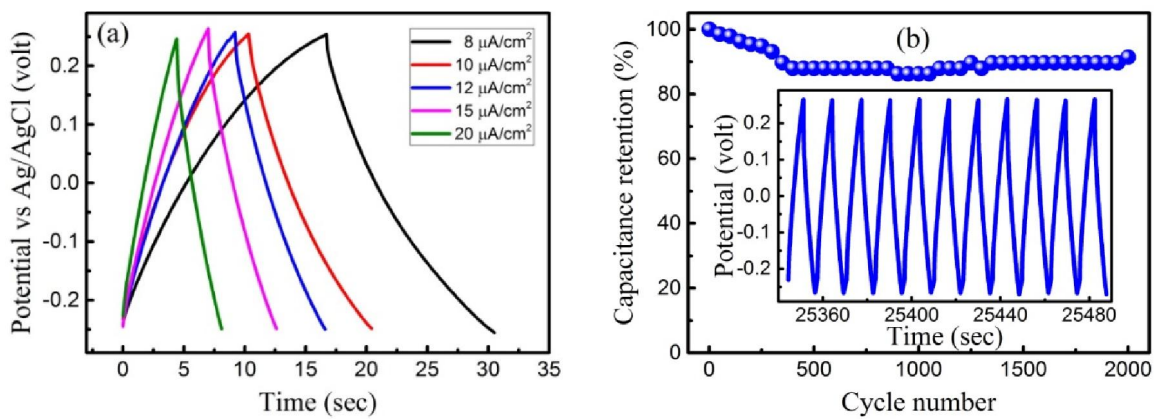

Fig. 6: (a) Charge-discharge profile of transferred VGNs at different current density of 8 – 20 µA/cm² and (b) capacitance retention versus cycle number, inset representative of few charge/discharge cycle.

The obtained experimental result supports the suitability of VGNs as promising electrode material, where open space of this novel nanoarchitech furnishes channel for the ion diffusion into the material. Furthermore, nanographitic base layer, from onset of the vertical network, provides mechanical support and serve as current collector. These findings envisioned the utility of VGNs as flexible supercapacitor electrode without the use of binder and additional current collector.

## 4. Conclusions

An efficient and easily scalable strategy for the transfer of vertical graphene nanosheets (VGNs) onto flexible substrate is presented. The preservation of morphology and structural quality of VGNs is confirmed by scanning electron microscopic and Raman spectroscopic analysis, respectively. The transfer process is very easy to implement and highly reproducible. Very low sheet resistance of the VGNs made them potential candidate for electrode materials. We have shown the utility of transferred VGNs as supercapacitor electrode without the use of any additional current collector. The present scalable transfer procedure holds great promise in conjunction with reliable and flexible electronic device applications, which is of great significance.

**Acknowledgement**

G.S would like to thanks Department of Atomic Energy, Govt. of India for Junior Research Fellowship. S.G. would like to acknowledge Department of Atomic Energy, Govt. of India for Postdoctoral Research Associateship. We would like to give thanks to Dr. Sandip Dhara for allowing Raman spectroscopy facility.

**Author contribution**

S.G. planned the work, assisted the experiments and wrote the manuscript. G.S. carried out the growth and transfer of vertical graphene nanosheets. S.G. performed the electrochemical experiment. All authors discussed the results, commented on the manuscript and gave approval to the final version of the manuscript.